# Magnetic characterization and switching of Co nano-rings in current-perpendicular-to-plane configuration

T. Yang [a)], M. Hara, and A. Hirohata

*Frontier Research System, RIKEN, 2-1 Hirosawa, Wako, Saitama 351-0198, Japan*

T. Kimura and Y. Otani

*Institute for Solid State Physics, University of Tokyo, Kashiwa, Chiba 227-8581, Japan,*

*and Frontier Research System, RIKEN, 2-1 Hirosawa, Wako, Saitama 351-0198, Japan*

**Abstract**

We fabricated Co nano-rings incorporated in the vertical pseudo-spin-valve nanopillar structures with deep submicron lateral sizes. It is shown that the current-perpendicular-to-plane giant magnetoresistance can be used to characterize a very small magnetic nano-ring effectively. Both the onion state and the flux-closure vortex state are observed. The Co nano-rings can be switched between the onion states as well as between onion and vortex states not only by the external field but also by the perpendicularly injected dc current.

a) E-mail:tyang@riken.jp

Nanomagnets have attracted considerable attention since they display different stable magnetic states depending on the shape, dimension, and composition.[1,2] Their potential technological applications include high-density magnetic random access memory (MRAM),[2] magnetic logic,[3] or other magnetoelectronic devices. Among the nanomagnets with various sizes and shapes, the magnetic rings have recently been conspicuous both theoretically [4] and experimentally,[5-8] attributed to their high symmetry and hence the very simple and well-defined magnetic states, i.e. the flux-closure vortex state and the onion state.[6-9] The magnetic ring is not only an ideal candidate for MRAM but also suited for investigating the nucleation, movement and annihilation of domain walls in well-controlled structures .

The magnetic switching in rings from submicron to micron scales has been investigated primarily with magneto-optical Kerr effect (MOKE) measurements, [6, 10, 11] compared with theoretical simulations. To provide sufficiently large signal, rings in an array with a large area are usually fabricated and their collective magnetic properties are measured. However, because of the sample-to-sample variation, the collective result does not necessarily reflect the behavior of an individual ring. To find the intrinsic switching properties, a single magnetic ring has been studied with probes like magnetic force microscopy (MFM),[7] photoemission electron microscopy (PEEM),[11] and Lorentz microscopy.[12] Magnetoresistance (MR) measurements have also been used to characterize a single ring in the current-in-plane (CIP) configuration.[9,13] Unfortunately, limited by the resolution or the size of the electrode, those probes are not suitable for a very small magnetic ring of sub-200 nm or smaller, which may present unique magnetic behaviors and is requested by such device applications as data storage to achieve high density. On the other hand, the giant magnetoresistance (GMR) measurement in the current-perpendicular-to-plane (CPP) configuration is not limited by the element size and thus suitable for characterizing a very small magnetic ring. In fact, for the CPP-GMR measurement, the smaller the element is, the stronger the signal is. The CPP-GMR is also easy to implement in a real magnetoelectronic device. In addition, the CPP configuration can be utilized to study the magnetization switching with a perpendicularly injected dc current.[14-18] The current-induced magnetization switching (CIMS) is a new approach to switch a nanomagnet through transferring spin angular momentum. CIMS can greatly simplify the structure of a magnetoelectronic device and reduce the power consumption. Recent studies have shown that the current in the CIP configuration induces the domain wall motion in the magnetic rings by applying a spin-transfer torque.[19, 20] A challenge in the CPP-GMR measurement is to fabricate the magnetic nanoring as small as sub-200 nm in a CPP configuration.

To study the magnetic nano-ring in the CPP configuration, we have fabricated vertical pseudo-spin-valve nanopillars from a magnetic multilayer Co20nm/Cu10nm/Co3nm/Au10nm, in which the Cu10nm/ Co3nm/Au10nm layers are patterned into the ring shape with various lateral sizes. Different from the lift-off process popularly used in fabricating the magnetic rings, ion milling is performed after patterning with electron beam lithography in our fabrication process, similar to that for an elliptical nanopillar. [21] The switching of the Co ring is thus studied with the CPP-GMR by applying either an in-plane external field $H_e$ or a vertical dc current. The resistance is measured with a small ac current and the lock-in technique at room temperature. The electrical current flowing from the bottom to the top is defined as positive, as depicted in Fig.1.

The SEM images of the fabricated rings are shown in Fig. 2. The widths of the rings are roughly about 50 nm, while the outer diameters are from 155 nm to 325 nm. The magnetic switches of the rings are firstly studied by sweeping the external field. The MR loops plotted in Fig.3 (a), (b), (c) and (d) correspond to the rings in Fig. 2 (a), (b), (c) and (d) respectively. As shown in Figs.3 (b), (c) and (d), three states are observed for the Co rings. According to the resistance values, they are respectively the onion state with the magnetization parallel to the bottom Co layer magnetization (hereafter onion-P state), the vortex state, and the onion state with the magnetization antiparallel to the bottom Co layer magnetization (hereafter onion-AP state). For the smallest Co ring, only two states are observed in Fig.3 (a). The one with smaller resistance is the onion-P state. The other state is characterized to be the onion-AP state from the minor MR loop showing that the switching of the bottom Co layer transforms it into the onion-P state. In addition, for the largest Co ring, the transition from the onion-AP state to the vortex state occurs at a very small external field. Therefore, in general, the vortex state becomes more and more unfavorable as the lateral size of the Co ring is reduced. This may be ascribed to the increasing curvature of the ring which results in an increased exchange energy according to the micromagnetic simulation.[8, 22]

To study the CIMS, the dc current is injected into the Co rings with the sizes of 155 nm/60 nm and 215 nm/95 nm. The dependences of the differential resistance d$V$/d$I$ on the dc current $I_{dc}$ are shown in Fig.4 (a) and (b) respectively. During the measurement, a small field of 40 Oe is applied to magnetize the extended bottom Co layer. According to the resistance levels, the two states appearing in the loop in Fig.4 (a) are onion-P and onion-AP states respectively. Therefore, a positive dc current transforms the onion-P state into the onion-AP state which is switched back to the onion-P state by a negative dc current. The switching behaviors for the smallest Co ring induced by both the external field and the dc current are very similar to those for the elliptical nanopillar studied previously.[21]

On the other hand, for the 215 nm/95 nm Co ring, two different d$V$/d$I$ ~ $I_{dc}$ loops are observed when the initial states are onion-AP and onion-P respectively, as drawn in Fig.4 (b). According to the resistance levels at 0 mA dc current, the magnetic transitions occur only between the initial onion state and a vortex state. For either loop, the favorable states are the high resistance state for positive currents and the low resistance state for negative currents, typical of the spin-transfer-induced magnetization switching. Nevertheless, to explain the switching behaviors, we should take into account not only the spin-transfer torque, but also the external field $H_e$ and the Oersted field $H_i$, which can be calculated with [23]

$$H_i = (I/2\pi R)(R^2 - R_{in}^2)/(R_{out}^2 - R_{in}^2), \tag{1}$$

where $R$ is the distance to the center of the ring, $R_{in}$ and $R_{out}$ are the inner and outer radii respectively. For a 10 mA dc current, $H_i$ is about 100 Oe at $R=(R_{out}+R_{in})/2$ in the ring. When a negative current is applied, the spin-transfer torque favors an P state while the circumferential Oersted field favors the clockwise vortex state, as shown in Fig.5 (a). For domain B in the onion-AP state, the Oersted field is in the same direction with the external field assisting the spin-transfer torque to reverse the magnetization; while for domain A, the Oersted field resists the spin torque. Consequently, only domain B is switched by the negative current, leading to the clockwise vortex state in Fig.5 (b).

When the current is reversed, both the Oersted field and the spin torque applied on the clockwise vortex state are also reversed, inducing domain nucleation in B and finally the transion to the onion-AP state, as shown in Fig.5 (c) and (d).

The switching between the onion-P state and the vortex state may be explained with a similar mechanism. However, the vortex evolved from the onion-P state should be counterclockwise. The resistance dependence on the current for the clockwise vortex state is different from that for the counterclockwise vortex state, as shown in Fig.4 (b).

It was reported[20] that the critical current density to move the domain wall in the ring is in the order of magnitude of $10^{12}$ A/m$^2$ in the CIP configuration. The critical switching current densities for the 215nm/95nm Co nano-ring are $2.5 \times 10^{11}$A/m$^2$, $3.3 \times 10^{11}$A/m$^2$, $3.1 \times 10^{11}$A/m$^2$, and $4.1 \times 10^{11}$A/m$^2$ for Onion-P to vortex, vortex to Onion-P, Onion-AP to vortex, and vortex to Onion-AP switches respectively, the same level as that for reversing the elliptical nanopillar.[21] It is also noticed from both Fig.3 (b) and Fig.4 (b) that the transition from the vortex to the onion is more difficult than that from the onion to the vortex. That may be due to the domain nucleation during the transition from the vortex to the onion states. Another explanation is that the vortex is more stable.

The 155nm/60nm ring does not show the vortex state during its switching. There are also two possible explanations. One is that the onion state may be more stable. The other is that when the size is reduced, the Oersted field may be negligible, thus the onion state is switched directly to the opposite onion state by the spin-transfer torque.

In summary, we have fabricated the ring-shaped nanopillars with the size down to sub-200 nm. Both the onion state and the vortex state in the Co nano-ring can be characterized with the CPP-GMR measurement. Perpendicularly injected dc current can induce magnetic transitions between onion and onion states as well as between onion and vortex states, attributed to the spin-transfer torque assisted by the Oersted field. In addition, the chirality of the vortex state is deduced from the direction of the current.

The authors are grateful to Dr. Tsukagoshi and the Nanoscience Development and Support Team of RIKEN for their great supports.

**References**

[1] C. Stamm, F. Marty, A. Vaterlaus, V. Weich, S. Egger, U. Maier, U. Ramsperger, H. Fuhrmann, and D. Pescia, Science **282**, 449 (1998).

[2] S. A. Wolf, D. D. Awschalom, R. A. Buhrman, J. M. Daughton, S. von Molnár, M. L. Roukes, A. Y. Chtchelkanova, and D. M. Treger, Science **294**, 1488 (2001).

[3] R. P. Cowburn and M. E. Welland, Science **287**, 1466 (2000).

[4] J. G. Zhu, Y. Zheng, and G. A. Prinz, J. Appl. Phys. **87**, 6668 (2000).

[5] K. Bussmann, G. A. Prinz, R. Bass, J.G. Zhu, Appl. Phys. Lett. **78**, 2029 (2001).

[6] J. Rothman, M. Klaui, L. Lopez-Diaz, C. A. F. Vaz, A. Bleloch, J. A. C. Bland, Z. Cui, and R. Speaks, Phys. Rev. Lett. **86**, 1098 (2001).

[7] S. P. Li, D. Peyrade, M. Natali, A. Lebib, Y. Chen, U. Ebels, L. D. Buda, and K. Ounadjela, Phys. Rev. Lett. **86**, 1102 (2001).

[8] M. Kläui, J. Rothman, L. Lopez-Diaz, C. A. F. Vaz, J. A. C. Bland, and Z. Cui, Appl. Phys. Lett. **78**, 3268 (2001).

[9] M. Kläui , C. A. F. Vaz, J. A. C. Bland, W. Wernsdorfer, G. Faini, E. Cambril, Appl. Phys. Lett. **81**, 108(2002).

[10] M. Kläui, C. A. F. Vaz, L. Lopez-Diaz, J. A. C. Bland, J. Phys. Condens. Matter **15**, 985(R) (2003).

[11] M. Kläui , C. A. F. Vaz, J. A. C. Bland, T. L. Monchesky, J. Unguris, E. Bauer, S. Cherifi, S.

Heun, A. Locatelli, L. J. Heyderman, Z. Cui, Phys. Rev. B **68**, 134426 (2003).

[12] J. N. Chapman and M. R. Scheinfein, J. Magn, Magn. Mater. **200**, 729 (1999).

[13] F. J. Castaño, D. Morecroft, W. Jung, and C. A. Ross, Phys. Rev. Lett. **95**, 137201 (2005).

[14] J. C. Slonczewski, J. Magn. Magn. Mater. **159**, L1 (1996).

[15] L. Berger, Phys. Rev. B **54**, 9353(1996).

[16] M. Tsoi, A. G. M. Jansen, J. Bass, W.-C. Chiang, M. Seck, V. Tsoi, and P. Wyder, Phys. Rev. Lett. **80**, 4281(1998).

[17] J. A. Katine, F. J. Albert, R. A. Buhrman, E. B. Myers, and D. C. Ralph, Phys. Rev. Lett. **84**, 3149(2000).

[18] Y. Jiang, S. Abe, T. Ochiai, T. Nozaki, A. Hirohata, N. Tezuka, and K. Inomata, Phys. Rev. Lett. **92**, 167204 (2004).

[19] Z. Q. Lu, Y. Zhou, Y. Q. Du, R. Moate, D. Wilton, G. H. Pan, Y. F. Chen, and Z. Cui, Appl. Phys. Lett. **88**, 142507 (2006).

[20] M. Laufenberg, W. Bührer, D. Bedau, P.-E. Melchy, M. Kläui, L. Vila, G. Faini, C. A. F. Vaz, J. A. C. Bland, and U. Rüdiger, Phys. Rev. Lett. **97**, 046602 (2006).

[21] T. Yang, A. Hirohata, T. Kimura, and Y. Otani, J. Appl. Phys. **99**, 073708 (2006).

[22] L. J. Heyderman, C. David, M. Kläui, C. A. F. Vaz, and J. A. C. Bland, J. Appl. Phys. **93**, 10011 (2003).

[23] J.Guo and M. B. A. Jalil, IEE Trans. Magn. **40**, 2122 (2004).

**Figure captions**

Figure 1 Schematic structure and measurement layout of the ring-shaped magnetic nanopillar.

Figure 2 SEM images for nano-rings with four different lateral sizes. The first row is the top view while the second row is the oblique side view. The outer/inner diameters in nm are labeled below the SEM images.

Figure 3 The CPP-GMR loops for nanopillars shown in Fig.2. The dashed line in (a) is the minor loop.

Figure 4 Current-induced magnetization switching loops for ring-shaped nanopillars with the sizes of (a) 155nm/60nm and (b) 215nm/95nm respectively. The dotted and solid lines in (b) are the loops measured with onion-AP and onion-P initial states respectively, as indicated by the solid circles. The open circle indicate the vortex state. To show the details, the loops are not drawn for the whole sweeping current range from +25 mA to -25 mA.

Figure 5 Illustration of the switching mechanisms for the magnetic transitions between AP-onion and vortex states. $M_1$ and $M_2$ (solid lines) are the magnetizations of the bottom Co layer and the Co nano-ring respectively. The dotted and dashed lines indicate the Oersted field and the dc current respectively.

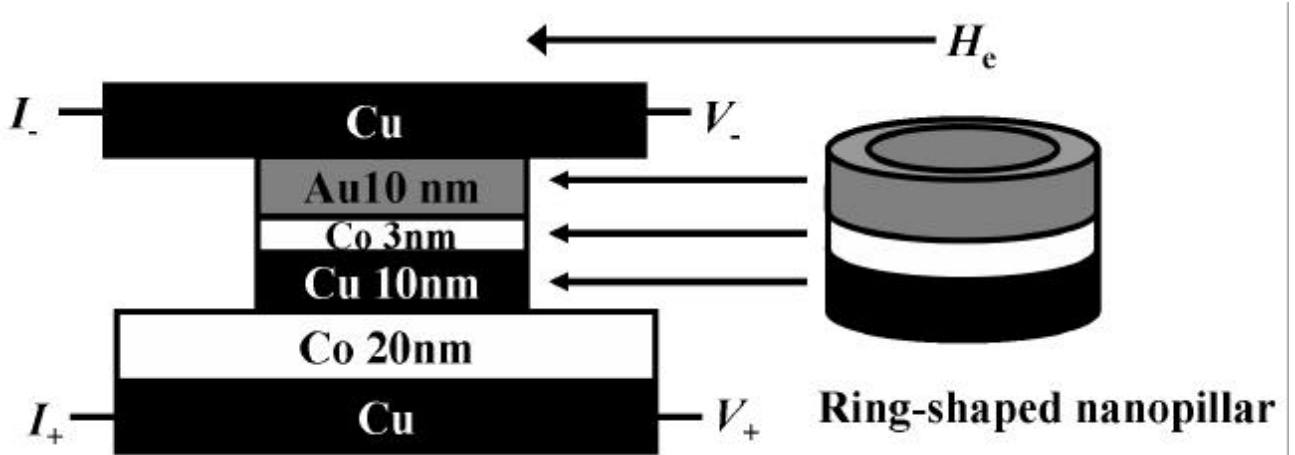

Figure 1

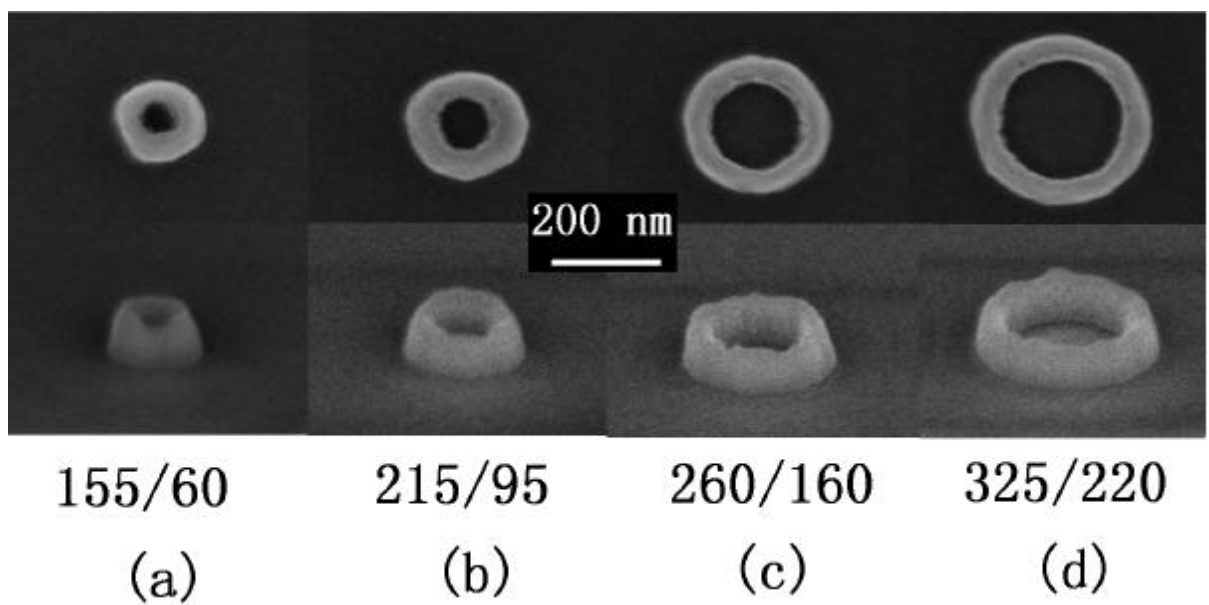

155/60  215/95  260/160  325/220
(a)    (b)    (c)    (d)

Figure 2

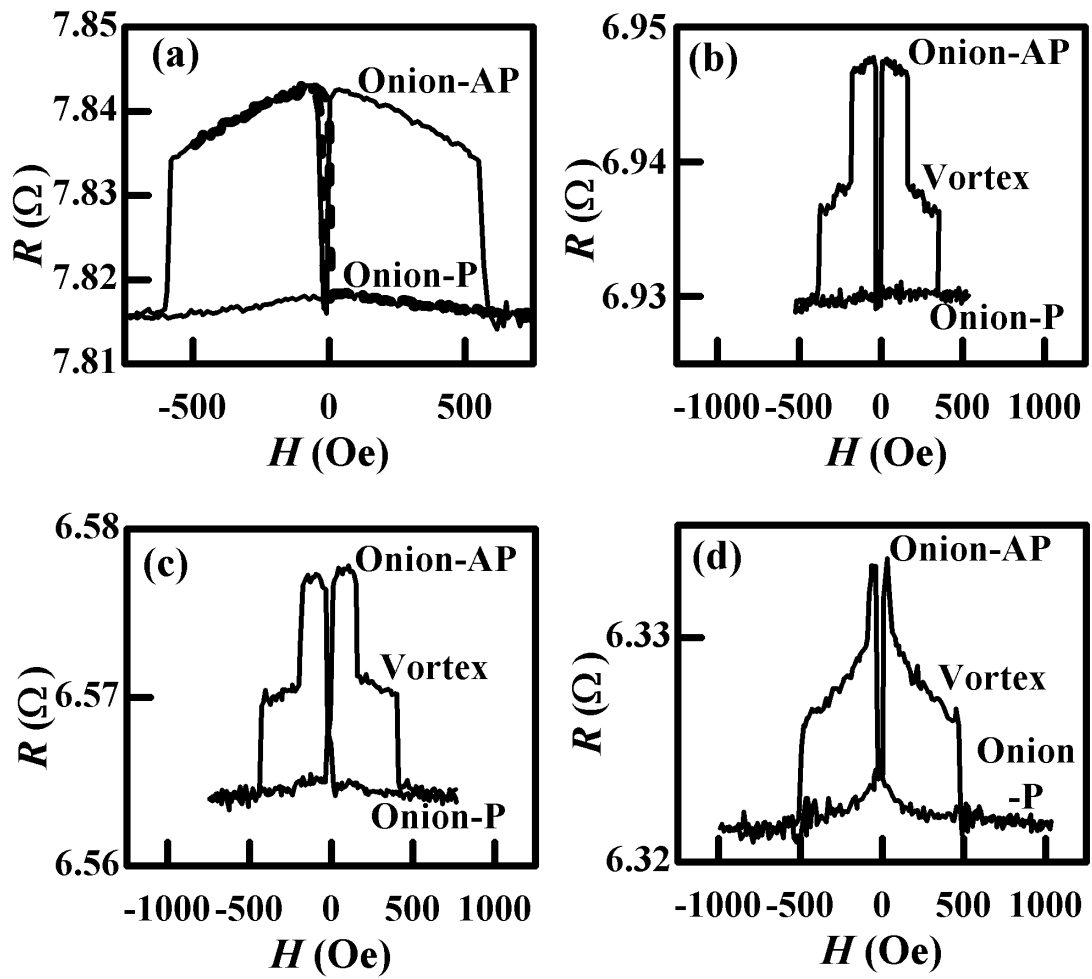

Figure 3

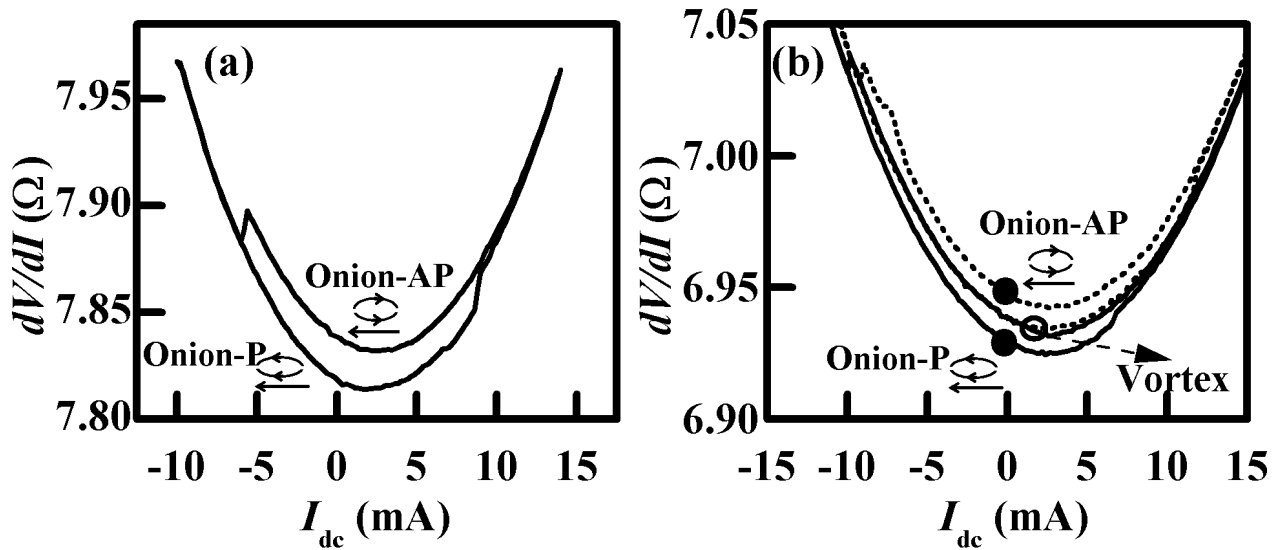

Figure 4

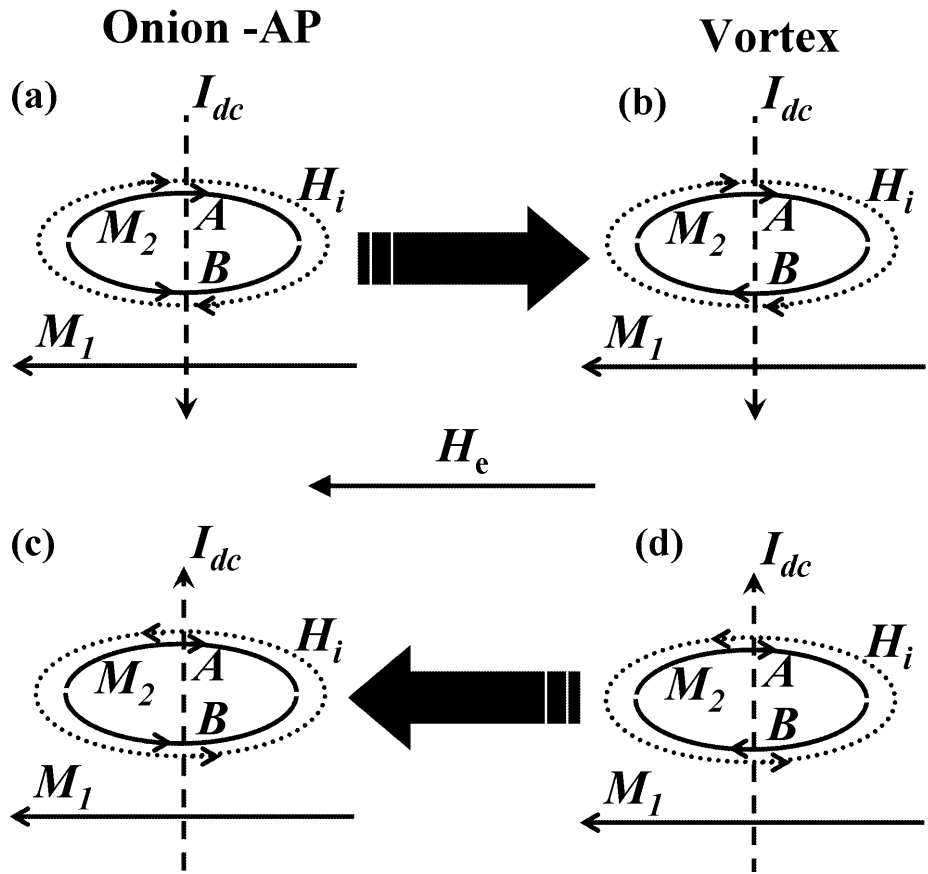

Figure 5